\documentclass[12pt]{article}
\usepackage{epsfig,amssymb,amsmath,psfrag,subfigure,rotate,color,wasysym}

\allowdisplaybreaks
\newcommand{\insertfig}[2]{\includegraphics[width=#1cm]{#2}}

\DeclareMathOperator*{\SumInt}{%
\mathchoice%
  {\ooalign{$\displaystyle\sum$\cr\hidewidth$\displaystyle\int$\hidewidth\cr}}
  {\ooalign{\raisebox{.14\height}{\scalebox{.7}{$\textstyle\sum$}}\cr\hidewidth$\textstyle\int$\hidewidth\cr}}
  {\ooalign{\raisebox{.2\height}{\scalebox{.6}{$\scriptstyle\sum$}}\cr$\scriptstyle\int$\cr}}
  {\ooalign{\raisebox{.2\height}{\scalebox{.6}{$\scriptstyle\sum$}}\cr$\scriptstyle\int$\cr}}
}

\def\XXint#1#2#3{{\setbox0=\hbox{$#1{#2#3}{\int}$ }
\vcenter{\hbox{$#2#3$ }}\kern-.6\wd0}}

\def \be  {\begin{equation}}
\def \ee  {\end{equation}}
\def \ba  {\begin{eqnarray}}
\def \ea  {\end{eqnarray}}
\def \baa {\begin{eqnarray*}}
\def \eaa {\end{eqnarray*}}
\def \lab #1 {\label{#1}}

\newcommand\re[1]{(\ref{#1})}
\def\d{\hbox{{d}\kern-.20em\hbox{l}}}

\def \matrix #1 {\left(\begin{array}{cc} #1 \end{array}\right)}

\def \tr {\mathop{\rm tr}\nolimits}

\newcommand \vev [1] {\langle{#1}\rangle}

\newcommand \ket [1] {|{#1}\rangle}
\newcommand \bra [1] {\langle {#1}|}

\def \ProdPrime {\mathop{\prod\nolimits'}\limits}

\newcommand{\ft}[2]{{\textstyle\frac{#1}{#2}}}
\numberwithin{equation}{section}
\begin{document}

\begin{titlepage}

\thispagestyle{empty}

\vspace*{3cm}

\centerline{\large \bf Spectral determinants for twist field correlators}
\vspace*{1cm}

\centerline{\sc A.V. Belitsky}

\vspace{10mm}

\centerline{\it Department of Physics, Arizona State University}
\centerline{\it Tempe, AZ 85287-1504, USA}

\vspace{2cm}

\centerline{\bf Abstract}

\vspace{5mm}

Twist fields were introduced a few decades ago as a quantum counterpart to classical kink configurations and disorder variables in low dimensional field theories. In 
recent years they received a new incarnation within the framework of geometric entropy and strong coupling limit of four-dimensional scattering amplitudes. In this 
paper, we study their two-point correlation functions in a free massless scalar theory, namely, twist--twist and twist--anti-twist correlators. In spite of the simplicity of
the model in question, the properties of the latter are far from being trivial. The problem is reduced, within the formalism of the path integral, to the study of spectral 
determinants on surfaces with conical points, which are then computed exactly making use of the zeta function regularization. We also provide an insight into 
twist correlators for a massive complex scalar by means of the Lifshitz-Krein trace formula.

\end{titlepage}

\setcounter{footnote} 0

\newpage




\section{Introduction}

Two-dimensional conformal field theory on a (say, hyperelliptic) Riemann surface $\mathcal{R}$ can be reformulated as a theory on a branched covering of the 
complex plane, where each brach point $z_i$ corresponds to an insertion of a conformal primary field $V (z_i)$ \cite{Dixon:1986qv,Knizhnik:1987xp,Bershadsky:1987jk},
known as the branch point twist field. The argument goes as follows \cite{Dixon:1986qv,Knizhnik:1987xp,Bershadsky:1987jk}. Consider a small vicinity of the branch 
point $z_j$ of order $N$ (with a cut emanating to infinity) on the covering of the original Riemann surface with $N$ consecutive sheets enumerated by $\ell = 1, \dots, N$ 
and parametrized by the single-valued coordinate $\zeta$,
\begin{align}
\zeta = \sqrt[N]{z - z_j}
\, .
\end{align}
When a point $z = z_j + \varepsilon {\rm e}^{i \varphi}$ is taken around $z_i$, i.e., from $\varphi = 0$ to $\varphi = 2 \pi$, the coordinate $\zeta$ moves from the 
$\ell$-th to $(\ell+1)$-st sheet. Let the field theory on each sheet $\ell$ be described by an action $S_{(\ell)} = S[X_{(\ell)}]$. The complex fields $X_{(\ell)}$ belonging to 
different sheets do not interact with each other. We identify them across the cuts $[z_i, \infty)$,
\begin{align}
X_{(\ell)} (z) |_{z \in [z_j, \infty)} = X_{(\ell + 1)} (z) |_{z \in [z_j, \infty)}
\, ,
\end{align}
and obviously cyclically identify $N+1 = 1$. This corresponds to gluing sheets together into the original Riemann surface. The Boltzmann weight ${\rm e}^{- S}$ in 
the partition function is then given by the sum of $N$ copies of the model on a single complex plane $S = \sum_{\ell = 1}^N S_{(\ell)}$ and enjoys $Z_N$ permutation 
symmetry $\sigma X_{(\ell)} = X_{(\ell + 1)}$. The latter can be diagonalized by forming a linear combination of fields of the $N$ copies
\begin{align}
X_{[k]} = \sum_{\ell = 1}^N {\rm e}^{2 \pi i \frac{k \ell}{N}} X_{(\ell)} 
\, , \qquad
\sigma X_{[k]} = {\rm e}^{2 \pi i \frac{k}{N}}  X_{[k]} 
\, .
\end{align}
The action can be equivalently written in terms of $N$ (still decoupled copies of) these, i.e., $S = \sum_{k = 0}^{N - 1} S [X_{[k]}]$. The above equation immediately 
implies that the new fields $X_{[k]}$ are not single-valued but acquire a phase as their argument crosses the cuts
\begin{align}
X_{[k]} (z_{\circlearrowleft_j}) |_{z \in [z_j, \infty)} = {\rm e}^{2 \pi i \frac{k}{N}} X_{[k]} (z) |_{z \in [z_j, \infty)}
\, .
\end{align}
This information can be encoded into the operator product expansion of the field $X_{[k]} (z)$ with a twist field $V_{k/N} (z_j)$, such that
\begin{align}
X_{[k]} (z) V_{k/N} (z_j) = (z - z_j )^{k/N} :X_{[k]} V_{k/N}: (z) + \dots
\, .
\end{align}
For the entire $\mathcal{R}$, the integrand in the path integral is given by the product of Boltzmann weights ${\rm e}^{-S} = \prod_{k = 0}^{N - 1} 
{\rm e}^{- S [X_{[k]}]}$ and one introduces the corresponding branch point twist field by forming the product of twist fields over all $N$ copies
\begin{align}
\label{TwistToBranchPointTwist}
V (z_j) = \prod_{k = 1}^{N - 1} V_{k/N} (z_j)
\, .
\end{align}
Here we accounted for the fact that $V_0 = 1$. Consequently, the field theory on the original Riemann surface is reformulated in terms of the $N$ decoupled
copies on the complex plane but with insertions of branch point twist fields
\begin{align}
\int [DX]_{\mathcal R} \, {\rm e}^{- S_{\mathcal R}}
=
\int \prod_{k = 0}^{N-1} \left( [DX_{[k]}] {\rm e}^{- S [X_{[k]}]} \right) \prod_j V (z_j) 
\equiv
\vev{\prod_j V (z_j) }
\, .
\end{align}
We implicitly imply that some (or all) of the fields can be branch point anti-twists $\bar{V} (z_j) = \prod_{k = 1}^{N - 1} V_{- k/N} (z_j)$.

The consideration that follows will be done having two particular applications of the branch point twist fields in mind. The first one is the calculation of von 
Neumann $S_{\rm vN}$ geometric (or entanglement) entropy \cite{Holzhey:1994we} for a one-dimensional observer residing on a single interval $A = [0,x]$ 
with the rest of the real line $B = \mathbb{R}\setminus A$ being unattainable,
\begin{align}
S_{\rm vN} = - \tr_{\rm A} \rho_{\rm A} \ln \rho_{\rm A}
\, , \qquad
\rho_{\rm A} = \tr_{\rm B} \ket{\psi} \bra{\psi}
\, .
\end{align}
Here $\ket{\psi}$ is a pure state of the quantum system on $\mathbb{R}$ and traces are taken over the degrees of freedom on the labelled intervals.
A standard technique for computation of $S_{\rm sV}$ is the replica trick \cite{Holzhey:1994we,Calabrese:2004eu}: introducing $N$ copies of the system, 
computing the Renyi entropy $S_{\rm R}$ and constructing its proper analytic continuation in $N$, allows one to compute its derivative
\begin{align}
\label{sVfromR}
S_{\rm R} (N) = \tr_A \rho_A^N
\, , \qquad
S_{\rm vN} = - \left. \frac{d S_{\rm R} (N)}{d N} \right|_{N \to 1} 
\, .
\end{align}
The Renyi entropy can be cast in a form of the path integral over a hyperelliptic Riemann surface, $\zeta^N = z (z - x)$, with a branch cut $[0,x]$. Then according to 
the previous discussion, it admits (up to an overall, possibly divergent, constant) the form of the correlation function of a branch point twist field at $z_{1, \mu} = (x, 0)$ 
and anti-twist at $z_{2, \mu} = (0, 0)$
\cite{Cardy:2007mb},
\begin{align}
\label{RenyiE}
S_{\rm R} (N) = c_N \vev{\bar{V} (0) V (x)} 
\, .
\end{align}

The second application pertains to the recent use of the pentagon form factor program \cite{Basso:2013vsa,Basso:2013aha,Belitsky:2014sla} for computation of $L$-particle 
scattering amplitudes $\mathcal{A}_L$ in four-dimensional maximally supersymmetric Yang-Mills theory. The formalism relies on a dual description of on-shell massless 
amplitudes in terms of expectation values of Wilson loops stretched on a null polygonal contour tracing the the path formed by the particles' momenta 
\cite{Alday:2007hr,Drummond:2007cf,Brandhuber:2007yx}. Dynamical information is then encoded through the physics of excitations $\psi$ 
propagating on the two-dimensional world-sheet ending on the four-dimensional boundary. The amplitudes admit the following schematic representation
\begin{align}
\mathcal{A}_L = \SumInt_{1,2,\dots L - 5} \bra{0} \widehat{\mathcal{P}}\ket{\psi_{L - 5}} 
{\rm e}^{- \tau_{L-5} E_{L-5} + i \sigma_{L-5} P_{L-5} + i \varphi_{L-5} m_{L-5}}
\dots 
\bra{\psi_2} \widehat{\mathcal{P}}\ket{\psi_1}
{\rm e}^{- \tau_1 E_1 + i \sigma_1 P_1 + i \varphi_1 m_1}
\bra{\psi_1} \widehat{\mathcal{P}}\ket{0}
\, ,
\end{align}
as a sum over an infinite number of the so-called flux-tube excitations and their bound states with external geometrical data $(\tau_j, \sigma_j, \varphi_j)$ included through 
the phase factors parametrized by total energy $E_j$, momentum $P_j$ and (twice the) helicity $m_j$ of the intermediate state $\ket{\psi_j}$. The physics of transitions 
between adjacent intermediate states is determined by matrix elements of pentagon operators $\widehat{\mathcal{P}}$. At strong 't Hooft coupling $g^2$, integrating out 
all heavy fermionic and gauge modes, the only propagating degrees of freedom that remain are the six nearly massless scalars. They are described by the O(6) nonlinear 
sigma model \cite{AldMal07}. In the ultraviolet regime, due to asymptotic freedom of the theory, the sigma model coupling vanishes and one ends up with a free theory 
of five massless real scalars. These propagate on a background with conical singularities at the space-time positions of pentagon operators $z_j = (\sigma_j, \tau_j)$ on a 
two-dimensional plane \cite{AldGaiMal09}. Each cone has an excess angle of $\pi/2$ as it corresponds to folding the pentagon on a time-space square \cite{Basso:2014jfa}. 
Thus, in the vicinity of each cone, we are dealing with $(1+\ft14)$-sheeted Riemann surface. As a consequence, the pentagon operators $\widehat{\mathcal{P}}$ take on the 
meaning of (fractional) branch point twist fields $V$. One subtle point that one has to keep in mind is that the introduction of twist fields requires operating with complex 
rather than real fields so one effectively doubles the number of degrees of freedom to construct a microscopic definition of pentagon operators. This is a common trick, 
successfully employed in many different circumstances, e.g., in the Ising model \cite{Schroer:1978sy}. By merely taking the square root at the end reduces this number 
in half. In this manner, the (square of the) scattering amplitudes as strong coupling is given by the $(L-5)$-point correlation function of branch point twist fields 
\cite{Basso:2014jfa,Belitsky:2015lzw,Bonini:2016knr}
\begin{align}
\mathcal{A}_L^2 \stackrel{g^2 \to \infty}{=}
\vev{V (z_{L - 5}) \dots V (z_1) V (0)}
\, .
\end{align} 
In what follows, we will focus on the simplest $L=6$ case, i.e., the hexagon.

The focus of our current study will be on two-point correlation functions in free theories of an elementary massless complex scalar field. Our subsequent presentation is 
organized as follows. In the next section, we recall the definition of twist fields within the formalism of the path integral. In Sect.~\ref{TwistTwistSection}, we compute the 
twist-twist and then, in Sect.~\ref{TwistAntiTwistSection} twist--anti-twist correlator in massless theories by reducing them to the calculation of spectral determinants 
regularized by means of the zeta function. While the twist-twist sector enjoys the conventional power-like scaling with the distance typical for theories with infinite
correlation length, the latter one acquires an additional logarithmic factor intrinsic to logarithmic conformal field theories \cite{Gurarie:1993xq}. Finally, we conclude
and provide a sketch for generalization of the current techniques to massive noninteracting theories as well as to multipoint correlators. Appendix contains information
on elementary mathematical aspects of $q$-series relevant to proper analysis of twist--anti-twist functions.

\section{Twist fields and their correlators}

The field theoretical definition of the twist field $V_\alpha$ was proposed in Refs.\ \cite{Sato:1978ht,Schroer:1978sy} within the framework of the two-dimensional Ising 
model as a continuum analogue of lattice (dis)order operators \cite{Kadanoff:1970kz}. They were further generalized within the formalism of the path integral to field models 
with scalars and fermions \cite{Marino:1981we,Schroer:1982iq}, where $V_\alpha$ was shown to be defined by non-polynomial and not manifestly local composite operators 
built up from elementary quantum fields, namely,
\begin{align}
\label{TwistFieldOperator}
V_\alpha (z)= \exp \left( 2 \pi i \alpha \int_{C_{[z,\infty]}} dz^\prime_\mu \varepsilon_{\mu\nu} j_\nu (z^\prime) \right)
\, .
\end{align}
Here the integral runs along an arbitrary contour $C_{[x,\infty]}$ from the position of the operator insertion to infinity with the integrand determined by the U(1) current, which 
for the complex scalar reads
\begin{align}
j_\mu =  \left( \partial_\mu \phi^\ast  \right) \phi - \phi^\ast \left( \partial_\mu \phi \right)
\, .
\end{align}
Its conservation implies the path independence of the definition \re{TwistFieldOperator},
\begin{align*}
\int_{C_{[z,\infty]}} dz^\prime_\mu \varepsilon_{\mu\nu} j_\nu (z^\prime) 
= 
\int_S d^2 z^\prime \partial_\mu j_\mu (z^\prime)
+
\int_{C'_{[z,\infty]}} dz^\prime_\mu \varepsilon_{\mu\nu} j_\nu (z^\prime)
\, ,
\end{align*}
where the first term vanishes upon the application of the Stokes theorem with $\partial S = C \cup (-C')$. In the above equation, it is assumed that $0 < \alpha < 1$, 
which is all one needs for calculation of branch point twist fields from $\alpha = k/N$ and $0 \leq k < N$.

\begin{figure}[t]
\begin{center}
\mbox{
\begin{picture}(0,140)(190,0)
\put(0,-270){\insertfig{20}{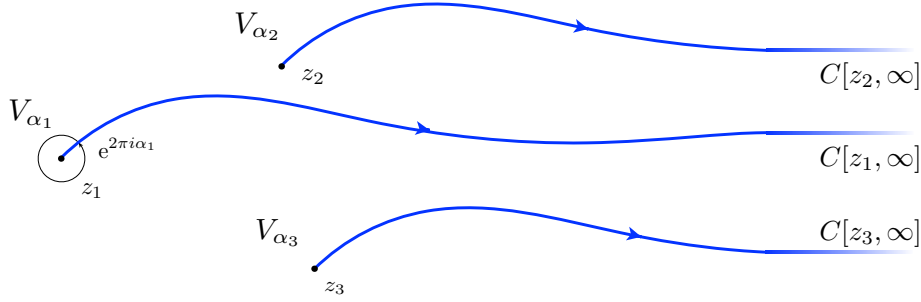}}
\end{picture}
}
\end{center}
\caption{ \label{FigTwistCorrelator} Generic correlation function of twist operators at insertion points $z_j$ and discontinuity contours $C [z_j, \infty]$,
with eigenfunctions of the corresponding Laplace equation developing twisted periodicity conditions \re{TwistedPeriods}. 
}
\end{figure}

As was demonstrated in Ref.\ \cite{Marino:1981we}, the correlation functions of twist fields in two-dimensional quantum field theory can be reformulated 
as statistical mechanics of matter fields in the external field of Dirac strings (or Aharonov-Bohm vortices after a gauge transformation). Namely, it is given by
the functional integral
\begin{align}
\label{TwistCorrFunct}
\vev{\prod_j V_{\alpha_j} (z_j)}
=
\left. \int [D \phi D \phi^\ast]
\, \exp \left( - \int d^2 z | D_\mu \phi (z) |^2 \right) \right/
\int [D \phi D \phi^\ast]
\, \exp \left( - \int d^2 z | \partial_\mu \phi (z) |^2 \right)
\, ,
\end{align}
where the gauge field in the covariant derivative $D_\mu = \partial_\mu - i A_\mu (z)$ is determined by the potential
\begin{align}
A_\mu (z)
=
2 \pi  \sum_j \alpha_j
\int_{[z_j, \infty)} d z^\prime_\nu \varepsilon_{\nu \mu} \delta^{(2)} (z^\prime - z)
\, .
\end{align}
See Fig.\ \ref{FigTwistCorrelator} for a graphical representation.

The path independence in the definition of the twist fields is traded off for the gauge invariance of the path integral. Then, making use of this freedom, we can 
choose all the paths as straight lines aligned with the $x$-axis. Without lost of generality, we place all operators along the $x$-axis, such that the vector potential is
\begin{align}
\label{DiracSrings}
A_x (z) = 0 \, , \qquad A_y (z) = 2 \pi \delta (z_y) \sum_i \alpha_i \theta (z_x - x_j) 
\, ,
\end{align}
where the components of any two-vector are $z_\mu = (z_x, z_y)$. Performing a gauge transformation, the external field takes on a form of a collection of 
the Aharonov-Bohm fluxes
\begin{align}
\label{ABvortex}
A_\mu (z) = \varepsilon_{\nu\mu} \sum_{j} \alpha_j \frac{(z - z_j)_\nu}{(z - z_j)^2}
\, .
\end{align}
We will not be using this form in the current paper though.

The resulting path integrals for the correlation function \re{TwistCorrFunct} being Gaussian in nature can immediately be evaluated and yield
\begin{align}
\vev{\prod_j V_{\alpha_j} (z_j)}
= 
\frac{\det \Delta_{0, 0, \dots} }{\det \Delta_{\alpha_1, \alpha_2, \dots} }
\, ,
\end{align}
with $\Delta_{\alpha_1, \alpha_2, \dots} = |D_\mu |^2$. The problem is thus reformulated as a spectral problem for the Laplace operator in the external field of
Dirac strings \re{DiracSrings}. Since the latter field exists only on the $x$-axis, one solves the free-space Laplace eigenvalue equation 
\begin{align}
\Delta_{0, 0, \dots} \Phi (z) = - E^2  \Phi (z)
\, , \qquad
z \notin \cup_j [z_j, \infty)
\, ,
\end{align}
with eigenfunctions possessing nontrivial monodromies around each point $z_j$,
\begin{align}
\label{TwistedPeriods}
\Phi (z_{\circlearrowleft_j} ) = {\rm e}^{2 \pi i \alpha_j} \Phi (z)
\, ,
\end{align}
with $z_{\circlearrowleft_j}$ denoting a $2 \pi$ rotation of $z = z_x + i z_y$ around $z_j$, see Fig.\ \ref{FigTwistCorrelator}.

Equivalently, the spectral problem can be viewed as the one for a Laplacian on Euclidean space with conical points $\{ z_j \}$. One peculiarity of this is that a choice 
has to be made in order to make it a self-adjoint operator. There are infinitely many possibilities to achieve it and they are driven by the prescription of a particular 
asymptotic behavior near the conical points from the functional space of the Laplacian. In the current paper, we will be considering Friedrichs extension, which, in 
physical terms, merely implies that its eigenfunctions are bounded near the conical points, see, e.g., Ref.\ \cite{Kokotov2009} for a comprehensive discussion. For 
the gauge transformed configuration \re{ABvortex}, it means that the Aharonov-Bohm vortices are impenetrable and, thus, the eigenfunctions of the Laplace operator 
vanish there. Let us point out that other self-adjoint extensions were discussed in Refs.\ \cite{KokotovHillairet,LMP2007}.

In order to properly calculate the determinants, we will employ the zeta regularization \cite{RaySinger,Hawking:1976ja} such that
\begin{align}
\det \Delta = \exp \left( - \mathcal{Z}^\prime (0) \right)
\, , 
\end{align}
and the zeta function
\begin{align}
\mathcal{Z} (s) = \sum_{n} E_n^{- 2 s}
\end{align}
sums up all strictly positive eigenvalues of the spectral problem
\begin{align}
\Delta \Phi_n = - E_n^2 \Phi_n
\, .
\end{align}
It is regular in the vicinity of $s = 0$. Recently, this technique was used to compute the vacuum expectation value of twist fields on a disk \cite{Belitsky:2017kcs} and 
explicit normalization coefficient naturally arises from that calculation. In the next two section, we extend the corresponding consideration to two-point functions.

\section{Massless twist--twist correlator}
\label{TwistTwistSection}

\begin{figure}[t]
\begin{center}
\mbox{
\begin{picture}(0,270)(225,0)
\put(0,-330){\insertfig{28}{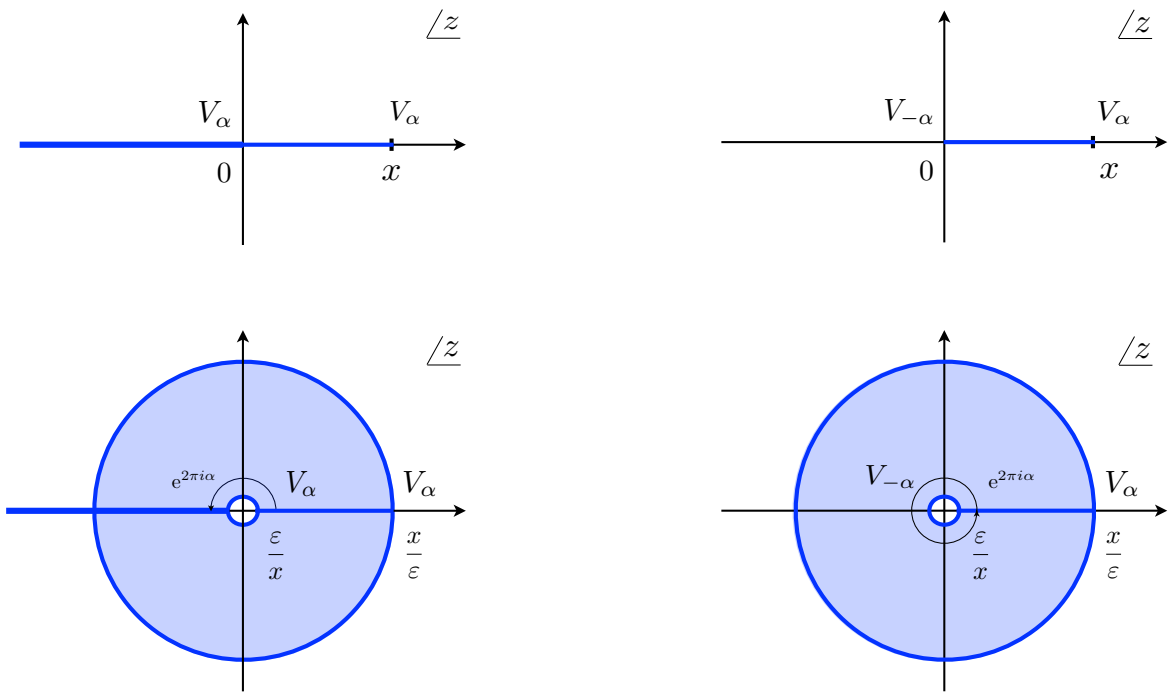}}
\end{picture}
}
\end{center}
\caption{ \label{MapFig} Cut structure from the insertion of twist--twist (left panel) and twist--anti-twist operators (right panel) in the complex plane.
Below each case is a graphical representation under the regularized form of the map \re{CoordinateMap}, when the twist operator at $z = x$ moves 
to infinity and explicitly exhibited twisted periodicity conditions. 
}
\end{figure}

Let us start our analysis with the two-point correlation function of twist operators in free massless theory. According to our previous discussion it is given by the
ratio of determinants of Laplace operators in free space and in the presence of two Dirac strings,
\begin{align}
\label{TwistTwistCorrelator}
\vev{V^{\rm s}_\alpha (0) V^{\rm s}_{\alpha} (x)} 
=
\frac{\det \Delta_{0, 0}}{\det \Delta_{\alpha, \alpha}} 
\, .
\end{align}
Making use of the path independence of twist operators, their contours are chosen as in Fig.\ \ref{MapFig} (top left panel). The coordinate transformation
\begin{align}
\label{CoordinateMap}
z_x \to \frac{z_x}{x - z_x}
\, ,
\end{align}
leaves the twist operator at $z_x = 0$ intact, while moving the one at $z_x = x$ to infinity. This map eliminate all scales from the problem, however. Therefore, one 
has to tread more carefully and introduce a regulator by cutting out an infinitesimal $\varepsilon$-vicinity around insertions before performing \re{CoordinateMap}, i.e., 
take $|z_x| \geq \varepsilon$ and $|z_x - x| \geq \varepsilon$. The resulting geometry takes on the form shown in Fig.\ \ref{MapFig} (bottom left panel).

To compute the determinant of the Laplacian $\Delta_{\alpha, \alpha}$, we have to solve the corresponding spectral problem. In polar coordinates, $z_x + i z_y
= r {\rm e}^{i \vartheta}$, the equation admits the form 
\begin{align}
\left( \frac{\partial^2}{\partial (\ln r)^2} + \frac{\partial^2}{\partial \vartheta^2}  \right)  \Phi (r, \vartheta) = - E^2 r^2 \Phi (r, \vartheta)
\, .
\end{align}
Separating the variables
\begin{align}
\Phi (r, \theta) = R (r) \Theta (\vartheta)
\, , \qquad
R (r) = J_{|\nu|} (E r)
\, , \qquad
\Theta (\vartheta) = {\rm e}^{i \nu \vartheta}
\, ,
\end{align}
the twisted periodicity condition on the angular function $\Theta (\pi) = {\rm e}^{2 \pi i \alpha} \Theta (0)$ provides the quantization condition for $\nu = 2 m + 2 
\alpha$ with integer $m = 0, \pm 1, \pm 2, \dots$. The requirement of the bounded behavior at the position of the twist fields fixes the radial wave function to the 
Bessel function of the first kind and yields the eigenvalues
\begin{align}
E_{m,n}^{(\alpha)} = j_{| 2m +2\alpha|, n}/(x/\varepsilon)
\, ,
\end{align}
in terms of the positive roots $j_{| 2m +2\alpha|, n}$ of $J_{|2m + 2 \alpha|}$. Making use of the zeta function regularization, we can rewrite this as
\begin{align}
\ln  \frac{\det \Delta_{0, 0}}{\det \Delta_{\alpha, \alpha}} 
=
\mathcal{Z}^\prime_1 (0, \alpha) 
+
\mathcal{Z}^\prime_2 (0, \alpha) 
+
\mathcal{Z}^\prime_2 (0, - \alpha)
- \mathcal{Z}^\prime_1 (0, 0) 
- 2 \mathcal{Z}^\prime_2 (0, 0)
\, ,
\end{align}
where the prime stands for the derivative with respect to the first argument and we introduced one- and two-dimensional Bessel zeta functions
\begin{align}
\label{Z1Z2}
\mathcal{Z}_1 (s, \alpha) = \sum_{n=1}^\infty \big( E_{0,n}^{(\alpha)} \big)^{- 2 s}
\, , \qquad
\mathcal{Z}_2 (s, \alpha) = \sum_{m=1}^\infty \sum_{n=1}^\infty \big( E_{m,n}^{(\alpha)} \big)^{- 2 s}
\, .
\end{align}

\subsection{One-dimensional contribution}

To start with, let us analyze the one-dimensional contribution. In fact, we do not even have to introduce $\mathcal{Z}_1$ as the determinant for $m=0$ case can be 
easily deduced from the application of the Gelfand-Yaglom theorem \cite{Gelfand} (see Ref.\ \cite{Dunne:2007rt} for a comprehensive introduction) since the problem 
is reduced to the one-dimensional radial problem. Let us start with this simple calculation and regularize the Laplacian with a mass $M$, $\Delta_{\alpha, \alpha} \to 
\Delta_{\alpha, \alpha} + M^2$, such that a regular solution at the origin $\Psi_\alpha (r) = J_{2 \alpha} (Mr)$ of the initial value problem
\begin{align}
\left( \Delta_{\alpha, \alpha} + M^2 \right) |_{m=0} \Psi_\alpha (r) = 0
\, ,
\end{align}
provides the value of the regularized determinant,
\begin{align}
\det \left( \Delta_{\alpha, \alpha} + M^2 \right) |_{m=0}  = \Psi_\alpha (x/\varepsilon)
\, .
\end{align}
Multiplying the right-hand side by the renormalization factor and taking the limit $M \to 0$, we immediately find
\begin{align}
\label{GYtheorem}
\frac{\det \Delta_0|_{m=0}}{\det \Delta_\alpha|_{m=0}} 
= 
\lim_{M \to 0} \frac{M^{2\alpha} \Psi_0 (x/\varepsilon)}{\Psi_\alpha (x/\varepsilon)} 
= 
2^{2 \alpha} \Gamma (1 + 2 \alpha) \left( \frac{\varepsilon}{x} \right)^{2 \alpha}
\, .
\end{align}

To verify this result, we now repeat it within the formalism of the zeta function regularization. To avoid dealing with the Bessel spectrum, $\mathcal{Z}_1$ is rewritten
in terms of the resolvent 
\begin{align}
\label{Z1}
\mathcal{Z}_1 (s, \alpha) = \int_{C^+} \frac{d \sigma}{2 \pi i} \sigma^{-s} \, \mathcal{R}_1 (\sigma, \alpha)
\end{align}
over a Hankel-like contour $C^+$ running counter-clockwise to the left of all poles of the integrand but to the right of the origin. Here $\mathcal{R}_1 (\sigma, \alpha) 
= - \mathcal{U}^\prime_1 (\sigma, \alpha)$  is traded for the potential $\mathcal{U}_1$
\begin{align}
\mathcal{U}_1 (\sigma, \alpha) 
=
- \ln \prod_{n=1}^\infty \left( 1 + \left( \frac{\sqrt{- \sigma}}{j_{2 \alpha, n}} \frac{x}{\varepsilon} \right)^2 \right)
=
- \ln I_{2 \alpha} \left( \frac{\sqrt{-\sigma} x}{\varepsilon} \right) + 2 \alpha \ln \frac{\sqrt{-\sigma} x}{2 \varepsilon} - \ln \Gamma (1 + 2 \alpha)
\, ,
\end{align}
and we employed the classic infinite series representation of the modified Bessel function $I_{2 \alpha}$ due to Watson \cite{Watson}. The function $\mathcal{U}_1$
vanishes at the origin $\sigma = 0$. It admits a systematic expansion in terms of powers and logarithms of $\sigma$ \cite{BruSee85,Spreafico}. In fact, the constant and 
$\log(-\sigma)$ term in its expansion determine the leading two terms in the small-$s$ expansion of the Bessel zeta function. Moving the derivative off $\mathcal{U}_1$
onto the function to its left and exponentiating $\sigma^{-s}$ by means of the Schwinger formula, we find
\begin{align}
\Gamma (s) \mathcal{Z}_1 (s, \alpha) = - s  \int_0^1 dt \, t^{s - 1}\int_{C^-} \frac{d \sigma}{2 \pi i} \frac{{\rm e}^{- t \sigma}}{\sigma} \, \mathcal{U}_1 (\sigma, \alpha)
+ O (s)
\, .
\end{align}
Here, we deformed the integration contour $C^+$ to the left of the origin and, since $\mathcal{U}_1 (0, \alpha) = 0$, the $C^-$ becomes the true Hankel contour. We also 
restricted the integration with respect to the Schwinger time over the half line to the interval $[0,1]$ which holds up to higher order powers of $s$. The large-$\sigma$
asymptotics of the potential reads
\begin{align}
\mathcal{U}_1 (\sigma, \alpha) = \mathcal{U}_{1,1} (\alpha)  \ln (- \sigma) + \mathcal{U}_{1,0} (\alpha) + \dots
\, , 
\end{align}
with
\begin{align}
\mathcal{U}_{1,1} (\alpha) = \frac{1}{4} + \alpha
\, , \quad
\mathcal{U}_{1,0} (\alpha) =  2 \left( \frac{1}{4} + \alpha \right) \ln \frac{x}{\varepsilon} - \ln \frac{2^{2 \alpha-1/2} \Gamma (1 + 2 \alpha) }{\sqrt{\pi}}
\, .
\end{align}
Evaluating the $\sigma$ and $t$ integrals, we find
\begin{align}
\Gamma (s) \mathcal{Z}_1 (s, \alpha) = \mathcal{U}_{1,1} (\alpha) \gamma_{\rm E} - \mathcal{U}_{1,0} (\alpha) - \frac{1}{s} \mathcal{U}_{1,1} (\alpha) + O (s)
\, ,
\end{align}
such that the expansion in $s$, yields
\begin{align}
\mathcal{Z}_1^\prime (0, \alpha) = - 2 \left( \frac{1}{4} + \alpha \right) \ln \frac{x}{\varepsilon} +  \ln \frac{2^{2 \alpha - 1/2} \Gamma (1 + 2 \alpha)}{\sqrt{\pi}}
\, ,
\end{align}
and coincides with Eq.\ \re{GYtheorem} obtained with the help of Gelfand-Yaglom theorem and mass regularization. 

\subsection{Two-dimensional contribution}
\label{Bessel2D}

Turning to the two-dimensional Bessel zeta function, we cast $\mathcal{Z}_2$ into the form
\begin{align}
\label{U2Potential}
\mathcal{Z}_2 (s, \alpha) 
= - s
\sum_{m=1}^\infty (m + \alpha)^{- 2 s} \int_{C^+} \frac{d \sigma}{2 \pi i} \, \sigma^{- s - 1} \, \mathcal{U}_2 (\sigma, m + \alpha) 
\, ,
\end{align}
with the potential
\begin{align}
\mathcal{U}_2 (\sigma, m + \alpha)
&
=
- \ln \prod_{n=1}^\infty \left( 1 + \left( \frac{(m + \alpha) \sqrt{- \sigma} }{j_{2 m + 2 \alpha, n}} \frac{x}{\varepsilon} \right)^2 \right)
\nonumber\\
&
= - \ln I_{2 m + 2 \alpha} \left( (m + \alpha) \frac{\sqrt{- \sigma} x}{\varepsilon} \right) + 2 (m + \alpha) \ln \frac{(m + \alpha) \sqrt{- \sigma} x}{2 \varepsilon}
- \ln \Gamma (1 + 2 m + 2 \alpha)
\, .
\end{align}

The calculation of the Kronecker limit formula for the double Bessel zeta function follows the footsteps of Ref.\ \cite{Belitsky:2017kcs} for twist operator on a disk. 
Since the sum accompanying the factor of $s$ in Eq.\ \re{U2Potential} is not analytic in $s$ in the vicinity of $s = 0$, we split $\mathcal{U}_2$ into two contributions
\begin{align}
\label{U2splitV2W2}
\mathcal{U}_2 (\sigma, m + \alpha) = \mathcal{V}_2 (\sigma, m + \alpha) + \frac{1}{m + \alpha} \mathcal{W}_2 (\sigma)
\, .
\end{align}
Notice that only the second term can generate a pole at $s=0$. The accompanying function $\mathcal{W}_2$ can be found from the uniform expansion of the Bessel 
function \cite{Olver} 
\begin{align*}
I_\nu (\nu z) 
\stackrel{\nu \to \infty}{=} 
\frac{{\rm e}^{\nu [ \sqrt{1 + z^2} + \ln z/(1 + \sqrt{1 + z^2}) ]}}{\sqrt{2 \pi \nu} (1 + z^2)^{1/4}} \left[ 1 + \frac{1}{\nu} U_1 (z) + O (\nu^{-2})\right]
\, ,
\end{align*}
where
\begin{align}
U_1 (z) = \frac{1}{8} (1 + z^2)^{- 1/2} - \frac{5}{24} (1 + z^2)^{-3/2}
\, ,
\end{align}
such that
\begin{align}
\mathcal{W}_2 (\sigma) = - \frac{1}{2} U_1 \left( \frac{\sqrt{-\sigma} x}{2 \varepsilon} \right)
\, .
\end{align}
The calculation of the integral associated with this contribution can be done immediately with the result
\begin{align}
\int_{C^+} \frac{d \sigma}{2 \pi i} \sigma^{- s - 1} \, \mathcal{W}_2 (\sigma) 
=
-  \frac{\Gamma (s + \ft12)}{24 \Gamma (s) \Gamma (\ft12)} \left( \frac{1}{s} + 5 \right) \left( \frac{x}{2 \varepsilon} \right)^{2s}
\, . 
\end{align}
The integral with $\mathcal{V}_2$ is evaluated analogously to the one-dimensional case, namely,
\begin{align}
\label{Bessel2DV2CplusToCminus}
\Gamma (s + 1) \int_{C^+} \frac{d \sigma}{2 \pi i} \sigma^{- s - 1} \mathcal{V}_2 (\sigma, m + \alpha)
=
&
- \mathcal{V}_2 (0, m + \alpha)
\\
&
-
\left( \gamma_{\rm E} - \frac{1}{s} \right) \mathcal{V}_{2,1} (m + \alpha) + \mathcal{V}_{2,0} (m + \alpha)
+ O (s)
\, , \nonumber
\end{align}
where $\mathcal{V}_{2,n}$ are extracted from the large-$\sigma$ expansion
\begin{align}
\label{LargeSigmaV2}
\mathcal{V}_2 (\sigma, m + \alpha) = \mathcal{V}_{2,1} (m + \alpha)  \ln (- \sigma) + \mathcal{V}_{2,0} (m + \alpha) + \dots
\, ,
\end{align}
with
\begin{align}
\mathcal{V}_{2,1} (m + \alpha)
&
=
\frac{1}{4} + m + \alpha
\, , \\ 
\mathcal{V}_{2,1} (m + \alpha)
&
=
2
\left(
\frac{1}{4} + m + \alpha
\right)
\ln \left( (m + \alpha) \frac{x}{\varepsilon} \right)
-
\frac{2^{2m + 2\alpha - 1/2} \Gamma (1 + 2m + 2\alpha)}{\sqrt{\pi}}
\, .
\end{align}
Putting these together, we obtain
\begin{align}
\Gamma (s)
\mathcal{Z}_2 (s, \alpha)
&
=
\sum_{m=1}^\infty (m + \alpha)^{- 2 s}
\left[
\mathcal{V}_2 (0, m + \alpha)
+
\left(
\gamma_{\rm E} - \frac{1}{s}
\right)
\mathcal{V}_{2,1} (m + \alpha)
-
\mathcal{V}_{2, 0} (m + \alpha)
\right]
\nonumber\\
&+
(1 + 5 s) \frac{\Gamma (s + \ft12)}{24 \Gamma (\ft12)} \left( \frac{x}{2 \varepsilon} \right)^{2 s} \sum_{m=1}^\infty (m + \alpha)^{- 2 s - 1}
+ O (s)
\, .
\end{align}
Performing the sums, the double zeta functions admits the form
\begin{align}
\Gamma (s)
\mathcal{Z}_2 (s, \alpha)
&
=
2 \zeta^\prime (2s - 1, 1 + \alpha) + \frac{1}{2} \zeta^\prime (2s, 1 + \alpha) 
-
\frac{1}{2} \ln (4 \pi) \zeta (2s, 1 + \alpha)
\nonumber\\
&
+
\left(
\gamma_{\rm E} - \frac{1}{s} - 2 \ln \frac{x}{2 \varepsilon}
\right)
[
\zeta (2s - 1, 1 + \alpha)
+
\ft{1}{4} 
\zeta (2s, 1 + \alpha)
]
\nonumber\\
&
+
\frac{\Gamma (s + \ft12)}{24 \Gamma (\ft12)} (1 + 5 s) \zeta (2s + 1, 1 + \alpha) \left( \frac{x}{2 \varepsilon} \right)^{2s} 
+ 
\chi_2 (s, \alpha) + O (s)
\, ,
\end{align}
in terms of the Hurwitz zeta functions and the function
\begin{align}
\chi_2 (s, \alpha) 
=
\sum_{m = 1}^\infty (m + \alpha)^{- 2s} \ln\Gamma (1 + 2m + 2\alpha) - \frac{1}{24} \zeta (2s + 1, 1 + \alpha)
\, ,
\end{align}
analytic at the origin $s = 0$. For the two-point twist-twist correlator \re{TwistTwistCorrelator}, we get
\begin{align}
\vev{V^{\rm s}_\alpha (0) V^{\rm s}_{\alpha} (x)} 
=
c^{\rm s}_{\alpha, \alpha} \left( \frac{\varepsilon}{x} \right)^{2 h_\alpha^{\rm s}}
\, ,
\end{align}
with the normalization constant being
\begin{align}
c^{\rm s}_{\alpha, \alpha} 
&
=
4^{\alpha (1 - \alpha)} \Gamma (1 + 2 \alpha) \exp \Big( \chi_2 (0,\alpha) + \chi_2 (0, - \alpha) -2 \chi_2 (0,0)  - \ft{1}{24} [\psi (1 - \alpha) + \psi (1 + \alpha) + 2 \gamma_{\rm E}] \Big) 
\, ,
\end{align}
and the well-known scaling dimension
\begin{align}
h_\alpha^{\rm s} = \alpha (1 - \alpha)
\, ,
\end{align}
of the scalar twist field \cite{Dixon:1986qv,Knizhnik:1987xp}.

\section{Massless twist--anti-twist correlator}
\label{TwistAntiTwistSection}

We turn now to the twist--anti-twist correlator, determined through the ratio
\begin{align}
\label{TantiTfromDet}
\vev{V^{\rm s}_{- \alpha} (0) V^{\rm s}_{\alpha} (x)} 
=
\frac{\det \Delta_{0, 0}}{\det \Delta_{-\alpha, \alpha}} 
\, .
\end{align}
Now, compared to just analyzed two-point function, the cut does not extend past the position of the (anti)twist fields, see the right panel in Fig.\ \ref{MapFig}. Therefore, 
one can use the conformal map $\varrho = \ln r$ from the annulus $\{ \varepsilon/x \leq r \leq x/\varepsilon, 0 \leq \vartheta \leq 2\pi \}$ to the cylinder $\{ -\varrho_0 \leq 
\varrho \leq \varrho_0, 0 \leq \vartheta \leq 2\pi\}$ with $\varrho_0 \equiv \ln (x/\varepsilon)$. It changes the metric by an overall factor $dr^2 + r^2 d\vartheta^2 = 
{\rm e}^{2 \varrho} (d \varrho^2 +  d\vartheta^2)$ and allows one to rewrite the determinant on the annulus in terms of the one on the cylinder making use of the Polyakov 
formula \cite{Polyakov:1981rd,Alvarez:1982zi,Osgood},
\begin{align}
\ln \det \Delta |_{\rm ann} = \ln \det \Delta |_{\rm cyl} - \frac{1}{3} \varrho_0
\, . 
\end{align}
The Jacobian however cancels between the vortex-dependent $\Delta_{- \alpha, \alpha}$ and the free $\Delta_{0, 0}$ Laplacians. The eigenspectrum equation on the 
cylinder is
\begin{align}
\left(
\frac{\partial^2}{\partial \varrho^2} + \frac{\partial^2}{\partial \vartheta^2}
\right)
\Phi (\varrho, \vartheta)
=
- E^2 \Phi (\varrho, \vartheta)
\, ,
\end{align}
with eigenfunctions in the separated variables being
\begin{align}
\Phi (\varrho, \vartheta) = R (\varrho) \Theta (\vartheta)
\, , \qquad
R (\varrho) = \sin \left( K (\varrho - \varrho_0) \right)
\, , \qquad
\Theta (\vartheta) = {\rm e}^{i \nu \vartheta}
\, ,
\end{align}
obeying the boundary conditions, $R (\varrho_0) = R (- \varrho_0) = 0$ and $\Theta (2 \pi) = {\rm e}^{2 \pi i \alpha} \Theta (0)$. The latter provide the quantization conditions
for the integration constants
\begin{align}
\nu = m + \alpha
\, , \qquad
K = \kappa n
\, , \qquad 
\kappa = \frac{\pi}{2 \varrho_0}
\, ,
\end{align}
with integer $0 < n$, $-\infty<m<\infty$ and omission of the $m=0$, for which the wave function vanishes identically. The eigenspectrum is thus
\begin{align}
\label{CylinderSpectrum}
E^{(\alpha)}_{n, m} = \sqrt{ (m + \alpha)^2 + (\kappa n)^2}
\, .
\end{align}
The correlation function of twist--anti-twist operators is then determined in its terms by the infinite product
\begin{align}
\vev{V^{\rm s}_{- \alpha} (0) V^{\rm s}_{\alpha} (x)} 
=
{\ProdPrime_{m = - \infty}^\infty} \prod_{n = 1}^\infty 
\left( \frac{E^{(0)}_{n, m}}{E^{(\alpha)}_{n, m}} \right)^2
\, .
\end{align}
The prime on the product stands for the omission of the ``zero mode" $m=0$.

\subsection{From twists to branch point twists}

Due to simplicity of the spectrum \re{CylinderSpectrum}, the resulting products can be analyzed directly without the use of the zeta function regularization. And one 
can deduce two different single-product representation which are amenable to further analysis. First, evaluating the product with respect to the index $n$ first, 
we obtain a representation 
\begin{align}
\label{VantiV1stRep}
\vev{V^{\rm s}_{- \alpha} (0) V^{\rm s}_{\alpha} (x)} 
=
{\ProdPrime_{m = - \infty}^\infty} \prod_{n = 1}^\infty \frac{(\kappa n)^2 + m^2}{(\kappa n)^2 + (m + \alpha)^2}
=
\frac{\sin (\pi \alpha)}{\pi \alpha}
\prod_{m=1}^\infty \frac{(1 - {\rm e}^{- 4 m \varrho_0})^2}{(1 - {\rm e}^{- 4 (m + \alpha) \varrho_0})(1 - {\rm e}^{- 4 (m - \alpha) \varrho_0})}
\, ,
\end{align}
where the products can be reexpressed in terms of Lambert's series. We will provide, however, an equivalent representation by first calculating the product in 
$m$ to make connection with previous analyses. Namely, rearranging the factors as
\begin{align}
{\ProdPrime_{m = - \infty}^\infty} \prod_{n = 1}^\infty \frac{(\kappa n)^2 + m^2}{(\kappa n)^2 + (m + \alpha)^2}
=
\prod_{n = 1}^\infty \prod_{m = 1}^\infty 
\left( 1 - \frac{\alpha^2}{(m - i \kappa n)^2} \right)^{-1} \left( 1 - \frac{\alpha^2}{(m + i \kappa n)^2} \right)^{-1}
\, , 
\end{align}
then making use of
\begin{align}
\prod_{m = 1}^\infty \left( 1 - \frac{\alpha^2}{(m + \kappa)^2} \right)
=
\frac{\Gamma^2 (1 + \kappa)}{\Gamma (1 - \alpha + \kappa) \Gamma (1 + \alpha + \kappa)}
\, ,
\end{align}
and eventually simplifying the product by means of $\Gamma (1 + Y) \Gamma (1 - Y) = (\pi Y)/\sin (\pi Y)$. These manipulations immediately yield
\begin{align}
\label{TwistAntitwistCorrelator}
\vev{V^{\rm s}_{- \alpha} (0) V^{\rm s}_{\alpha} (x)} = 
\prod_{n=1}^\infty \frac{(1 - q^n)^2}{(1 - {\rm e}^{2 \pi i \alpha} q^n)(1 - {\rm e}^{- 2 \pi i \alpha} q^n)}
\end{align}
where we introduced a compact notation
\begin{align}
q = {\rm e}^{2 \pi \kappa}
\, .
\end{align}

Since this correlator plays a distinguished role in defining the geometric entropy \cite{Cardy:2007mb} according to Eq.\ \re{RenyiE}, let us pass from the (anti)twist fields 
to branch-point (anti)twist operators. We use Eq.\ \re{TwistToBranchPointTwist} to find
\begin{align}
\vev{\bar{V}^{\rm s} (0) V^{\rm s} (x)} = \prod_{k = 1}^{N-1} \vev{V_{- k/N} (0) V_{k/N} (x)}
=
\left[ \prod_{n=1}^{\infty} \frac{(1 - q^n)^N}{(1 - q^{n N})} \right]^2
\, ,
\end{align}
where we made use of the fact that
\begin{align}
\prod_{k = 1}^{N - 1} (1 - Q {\rm e}^{\pm 2 \pi i k/N}) = \frac{1-Q^N}{1-Q}
\, .
\end{align}
This was the starting point of the analysis in Ref.\ \cite{Holzhey:1994we}. The infinite products in the above equation can be cast in terms of the well-studied
$q$-Pochhammer symbol, see, e.g., \cite{Andrews86},
\begin{align}
\prod_{n=1}^\infty (1 - Q^n) = (Q,Q)_\infty
\, ,
\end{align}
such that
\begin{align}
\vev{\bar{V}^{\rm s} (0) V^{\rm s} (x)} = \frac{(q,q)_\infty^{2N}}{(q^N, q^N)^2_\infty}
\end{align}
Then its ultraviolet asymptotics as $\varepsilon \to 0$ can be easily read off from the well-known behavior of the $q$-Pochhammer symbols as $q \to 1_-$ due to 
Watson estimation \cite{Kluyver19,WatsonEst} (see also Refs.\ \cite{Banerjee16} for recent accounts),
\begin{align}
\label{WatsonEstimation}
(q,q)_\infty = \left( \frac{2 \pi}{1 - q} \right)^{1/2} \exp\left( \frac{\zeta (2)}{\ln q} \right) [1 + O(q)]
\, .
\end{align}
Then the two-point function on an $N$-sheeted Riemann surface reads
\begin{align}
\vev{\bar{V}^{\rm s} (0) V^{\rm s} (x)} = N \left( \frac{\ln \frac{x}{\varepsilon}}{2 \pi} \right)^{N-1} \left( \frac{\varepsilon}{x}  \right)^{\frac{1}{6} \left(N - \frac{1}{N} \right)}
\, ,
\end{align}
and acquires a multiplicative logarithmic dependence, which are typical of logarithmic CFTs \cite{Gurarie:1993xq}. Making use of the definitions \re{sVfromR} and \re{RenyiE}, 
this fact implies the emergence of an additive $\ln\ln$-correction to the famous entanglement entropy scaling
\begin{align}
S_{\rm vN} \sim \frac{\rm c}{6} \ln\frac{x}{\varepsilon} - \ln\ln \frac{x}{\varepsilon} + {\rm const.}
\, ,
\end{align}
with ultraviolet cutoff $\varepsilon$ due to Ref.\ \cite{Holzhey:1994we}. Here $c = 2$ is the central charge of a complex scalar studied in this work. The second term 
in the above equation was recently noticed by numerically resumming form factor expansion to two-point functions in Ref.\ \cite{Bianchini:2016mra} and via analytical 
methods in Ref.\ \cite{Blondeau-Fournier:2016rtu}, though our sign is opposite to the one in the first study.

The asymptotic scaling of the correlation function \re{TwistAntitwistCorrelator} can be analyzed in a similar fashion. However, before we do it, in a general spirit of this 
paper, we will provide the derivation of \re{TwistAntitwistCorrelator} within the framework of zeta function regularization in the next section. This analysis is of interest in 
its own right as it provides a derivation of the Kronecker limit formula \cite{Weil76,Osgood} for the double spectral series in question.

\subsection{Zeta function regularization}

Let is calculate the two-point correlation function \re{TantiTfromDet} by means of the zeta function regularization. We introduce
\begin{align}
\mathcal{Z} (s, \alpha) = \sum_{n,m=1}^\infty [(m + \alpha)^2 + (\kappa n)^2]^{- s}
\, ,
\end{align}
which is a generalization of the Epstein zeta function \cite{Epstein1903}. As in Sect.~\ref{Bessel2D}, we write it as a contour integral 
\begin{align}
\mathcal{Z} (s,  \alpha) = - s \sum_{m=1}^\infty (m + \alpha)^{- 2 s} \int_{C_+} \frac{d \sigma}{2 \pi i} \sigma^{- s - 1} \mathcal{U} (\sigma, m + \alpha)
\, , 
\end{align}
in terms of the potential
\begin{align}
\mathcal{U} (\sigma, m + \alpha) 
&= 
- \ln \prod_{n = 1}^\infty \left( 1 + \frac{(m + \alpha)^2 (- \sigma)}{(m + \alpha)^2 + (\kappa n)^2} \right)
\nonumber\\
&=
\ln (1 - \sigma)
+
\ln \frac{
\Gamma \left( i \sqrt{1 - \sigma} (m + \alpha)/\kappa \right) \Gamma \left( - i \sqrt{1 - \sigma}  (m + \alpha)/\kappa \right)
}{
\Gamma \left( i (m + \alpha)/\kappa \right) \Gamma \left( - i  (m + \alpha)/\kappa \right)
}
\, .
\end{align}
First, one notices that $\mathcal{U} (0, m + \alpha) = 0$, so there is no contribution from the residue at $\sigma = 0$ in Eq.\ \re{Bessel2DV2CplusToCminus}.
Second, since the argument of the logarithm is an even function of $(m + \alpha)$, the application of the Stirling formula with the first power correction
accounted for, i.e., $\ln \Gamma (Y) = {\dots} + 1/(12 Y) + O(1/Y^3)$, implies that $\mathcal{W} = 0$ in the analogue of Eq.\ \re{U2splitV2W2} for the case at 
hand. One finds
\begin{align}
\Gamma (s + 1) \int_{C^+} \frac{d \sigma}{2 \pi i} \sigma^{- s - 1} \mathcal{U} (\sigma, m + \alpha)
=
-
\left( \gamma_{\rm E} - \frac{1}{s} \right) \mathcal{U}_{1} (m + \alpha) + \mathcal{U}_{0} (m + \alpha)
+ O (s)
\, , 
\end{align}
where the coefficients $\mathcal{U}_{1/0}$ accompanying $\ln^{1/0} (- \sigma)$ dependence in the asymptotic large-$\sigma$ expansion \re{LargeSigmaV2} 
are
\begin{align}
\mathcal{U}_{1} (m + \alpha)
&= \frac{1}{2}
\, , \\
\mathcal{U}_{0} (m + \alpha)
&=
\ln \left( 2 \sinh \frac{\pi (m + \alpha)}{\kappa} \right)
\, .
\end{align}
The Kronecker-like limit formula then reads
\begin{align}
\Gamma (s)
\mathcal{Z} (s, \alpha)
&
=
\sum_{m=1}^\infty (m + \alpha)^{- 2 s}
\left[
\left(
\gamma_{\rm E} - \frac{1}{s}
\right)
\mathcal{U}_{1} (m + \alpha)
-
\mathcal{U}_{0} (m + \alpha)
\right]
+ O (s)
\, .
\end{align}
In its terms, the correlation function is
\begin{align}
\ln \vev{V^{\rm s}_{- \alpha} (0) V^{\rm s}_{\alpha} (x)} 
=
\mathcal{Z}^\prime (0, \alpha) 
+
\mathcal{Z}^\prime (0, - \alpha)
- 
2 \mathcal{Z}^\prime (0, 0)
\, ,
\end{align}
and can immediately be found to coincide with Eq.\ \re{VantiV1stRep}.

Let us conclude this section with the ultraviolet asymptotics of twist--anti-twist correlator as $\varepsilon \to 0$. Making use of expansion formulas
as $q \to 1_-$ derived in Appendix \ref{AppendixAsympt}, we find the logarithmically enhanced behavior
\begin{align}
\label{TwisAntiTwistCorrAlpha}
\vev{V^{\rm s}_{-\alpha} (0) V^{\rm s}_{\alpha} (x)} 
=
c^{\rm s}_{-\alpha, \alpha} \left( \frac{\varepsilon}{x} \right)^{2 h_\alpha^{\rm s}} \ln \frac{x}{\varepsilon}
\, ,
\end{align}
with the normalization constant
\begin{align}
c^{\rm s}_{-\alpha, \alpha} = \frac{4 \sin (\pi \alpha)}{\pi} \, .
\end{align}
It is important to realize that the logarithms stems from the determinant of the free Laplacian $\Delta_{0,0}$, i.e., the numerator in Eq.\ \re{TantiTfromDet}, providing 
proper normalization for the correlation function in question. The result of Ref.\ \cite{Blondeau-Fournier:2016rtu}  obtained within the formalism of the angular 
quantization \cite{LukZam97} agrees with this expression.

\section{Instead of a conclusion}

In the main body of this paper, we focused on two-point twist--(anti)twist correlations for a massless free scalar. The massive case can be analyzed in a similar 
manner. To provide a sketch of the treatment, we will limit ourselves to the expectation value of twist field. The solutions $\Phi (r, \vartheta)$ to the eigenvalue 
equation with a singe twist, decaying at infinity as $\Phi (r \to \infty, \vartheta) \sim {\rm e}^{- M r}$, admits the form (when cast in a contour integral form)
\begin{align}
\label{DiskWaveFunction}
\Phi (r, \vartheta) = \int d\kappa \, c (\kappa) K_{i \kappa} (r) {\rm e}^{- \kappa \vartheta}
\, .
\end{align}
It endows the system with a discrete spectrum by imposing quasiperiodicity conditions on the angular dependence and the Dirichlet boundary condition $\Phi (r = \varepsilon, 
\vartheta) = 0$ on the circular domain $r = \varepsilon$ in the vicinity of the vortex $V_\alpha (0)$. The vacuum average of the scalar twist field then reads
\begin{align}
\ln \vev{V^{\rm s}_\alpha}_{M} 
= 
\sum_{n = 1}^\infty \ln\frac{(E^{(0)}_{0,n})^2 + M^2}{(E^{(\alpha)}_{0,n})^2 + M^2}
+
\sum_{m = 1}^\infty \sum_{n = 1}^\infty \ln \frac{(E^{(0)}_{m,n})^2 + M^2}{(E^{(\alpha)}_{m,n})^2 + M^2}\frac{(E^{(0)}_{m,n})^2 + M^2}{(E^{(- \alpha)}_{m,n})^2 + M^2}
\, ,
\end{align}
where as in the massless case, we separated the $m = 0$ eigenvalue. According to Ref.\ \cite{Marino:1981we}, the second term involving the double sum can be
related to the determinant of the Dirac operator with the Atiyah-Patodi-Singer spectral boundary conditions \cite{APSbcs,Hortacsu80,Marino:1981we} and, bosonising 
the fermion, it can then be reduced to the analysis of the vacuum average of a vertex operator in the dual picture of the massive sine-Gordon model \cite{LukZam97}. 
Presently, we will not address this term within the zeta function regularization, but the corresponding inhomogeneous double spectral series is of interest in its own right 
and requires a separate study. Here, we will merely provide the result for the single spectral series associated with $m = 0$ term since it is responsible for logarithmic
dependence of the vacuum expectation value in ultraviolet cutoff $\varepsilon$. The use of the Gelfand-Yaglom theorem immediately yields the following expression
\begin{align}
\prod_{n = 1}^\infty \frac{(E^{(0)}_{0,n})^2 + M^2}{(E^{(\alpha)}_{0,n})^2 + M^2}
=
\frac{K_0 (M \varepsilon)}{K_\alpha (M \varepsilon)}
\stackrel{\varepsilon \to 0}{\simeq}
\frac{(M \varepsilon)^\alpha}{2^{\alpha - 1} \Gamma (\alpha)} \ln \left( \frac{2 {\rm e}^{- \gamma_{\rm E}}}{M \varepsilon} \right)
\, .
\end{align}
In complete analogy to the massless case, discussed in Ref.\ \cite{Belitsky:2017kcs}, the double product develops $(M \varepsilon)^{- \alpha^2}$ dependence on 
ultraviolet cutoff completing the exponent of the $M$-dependence to the conformal dimension of the scalar twist field $h_\alpha$. The resulting functional dependence 
on the parameter $M \varepsilon$ is in agreement with previous studies, where the logarithmic corrections due to conformal symmetry breaking by mass perturbations 
were observed from analyses of Painlev\'e equation \cite{Sato:1978ht,McCoy:1976cd,Casini:2005zv}, form factor resummation \cite{Bianchini:2016mra} and angular 
quantization \cite{Blondeau-Fournier:2016rtu}.

The ratio of determinants defining correlators in question, also known as the relative spectral determinants \cite{Mueller96}, can be equivalently (and more efficiently) 
studied in noncompact space by means of the Lifshitz-Krein trace formula \cite{Lif52,Kre53}. For a function $f$ of a self-adjoint operator $A_0$ and its perturbation $A$, with 
their difference $A - A_0$ being of the trace class,
\begin{align}
\tr \left[ f (A) - f (A_0) \right] = \int d \kappa f (E_\kappa) \partial_\kappa \xi (\kappa)
\, .
\end{align}
Here, the spectral shift function $\xi (E)$, associated with the pair $(A_0,A)$, is given by the Birman-Krein formula \cite{BirKre62}
\begin{align}
\xi (\kappa) = \frac{1}{2 \pi i} \ln \det S (\kappa)
\, ,
\end{align}
in terms of the scattering matrix $S (E)$ on the perturbation $A - A_0$ (for a review, see \cite{BirYaf92}). Then, taking the function $f$ in the form of the heat kernel
$f (A) = \exp (- t A)$, the vacuum expectation is given by the relative determinant regularized by means of the zeta function \cite{Mueller96} 
\begin{align}
\label{MassiveVEV}
\ln \vev{V_\alpha (0)} = - \mathcal{Z}^\prime (0, \Delta_\alpha, \Delta_0)
\, ,
\end{align}
where
\begin{align}
\label{NoncompactZetaToS}
\mathcal{Z} (s, \Delta_\alpha, \Delta_0) = \int \frac{d \kappa}{2 \pi i} E_\kappa^{-s} \partial_\kappa \ln \det S (\kappa)
\, .
\end{align}

The scattering matrix on one vortex can be easily found from the asymptotic behavior of the eigenfunction \re{DiskWaveFunction} in the vicinity of the vortex
$r = 0$ as a ratio of coefficient between left and right movers,
\begin{align}
\Phi (r \to 0, \vartheta) \sim (r/2)^{- i \kappa} + S (\kappa) (r/2)^{i \kappa}
\, ,
\end{align}
with 
\begin{align}
S (\kappa) = \frac{\Gamma (1 - i \kappa)}{\Gamma (1 + i \kappa)}
\, .
\end{align}
With this input and explicit form of the resummed energy spectrum, see Eq.\ \re{TwistAntitwistCorrelator} above, one can immediately see that the expectation
value \re{MassiveVEV} is the same as obtained with the formalism of the angular quantization advocated in Ref.\ \cite{LukZam97} for fermions and recently 
applied to scalars in Ref.\ \cite{Blondeau-Fournier:2016rtu}. 

In light of the representation \re{NoncompactZetaToS}, one it is instructive to explore a connection to scattering matrices on Aharonov-Bohm vortices \re{ABvortex} 
as a tool to compute multi-twist correlators. Recently, scattering amplitudes on two fluxes were found explicitly as a solution to Painlev\'e V equations \cite{Bogomolny}. 
This formulation allows for an extension to systems with multiple vortices, i.e, surfaces with multiple conical singularities, \cite{KokotovHillairet} and it provides a natural 
geometric interpretation to the gluing of conical manifolds \cite{LMP2007} via analytic surgery procedure of Burghelea, Friedlander and Kappeler \cite{BFK1992}
by interpreting the S-matrix as a kind of limiting Dirichlet-to-Neumann operator. The application of these considerations to the current setup of correlation functions
of twist operators will be discussed elsewhere.

\section*{Acknowledgments}

This work was completed during the author's visit at ENS (Paris) and IPhT (Saclay). We would like to thank Benjamin Basso and Gregory Korchemsky and for the warm 
hospitality at respective institutions, and Leonid Friedlander, Ivan Kostov, Didina Serban and Frank Wilczek for discussions. This research was supported by the U.S. 
National Science Foundation under the grant PHY-1403891.

\appendix

\section{Asymptotic expansion}
\label{AppendixAsympt}

Let us calculate the asymptotic expansion of the product
\begin{align}
\prod_{k = 1}^\infty (1 - a q^k) = \frac{(a, q)_\infty}{1 - a}
\, ,
\end{align}
contributing to the cylinder partition function in the $q \to 1_-$ limit and $a \neq 1$. This can be done by elementary methods. The first step consists in exponentiating 
of the terms in the product and expanding the logarithm in the region of its convergence in the Taylor expansion and then exchanging the two sums. This yields
\begin{align}
\prod_{k = 1}^\infty (1 - a q^k) = \exp \left( \sum_{\ell = 1}^\infty \frac{1}{\ell} \frac{a^\ell}{1 - q^{ - \ell}} \right)
\, .
\end{align}
Then adopting the second representation, the $q \to 1_-$ expansion can be easily performed with the following chain of transformations
\begin{align}
\label{GenericaExpantion}
\ln \prod_{k = 1}^\infty (1 - a q^k) 
= 
\sum_{\ell = 1}^\infty  \frac{a^\ell}{\ell} \left( 1 - \sum_{n = 0}^\infty \frac{\ell^n}{n!} \ln^n \frac{1}{q} \right)^{-1}
=
- \frac{1}{\ln \frac{1}{q}} {\rm Li}_2 (a) - \frac{1}{2} \ln (1 - a) + O \left( \ln \frac{1}{q} \right)
\, .
\nonumber
\end{align}
The second factor in the correlation function \re{TwistAntitwistCorrelator} can be obtained by the inversion $a \to 1/a$. Thus, combining them together and making 
use of the obvious relations
\begin{align}
&
{\rm Li}_2 ({\rm e}^{2 \pi i \alpha}) + {\rm Li}_2 ({\rm e}^{- 2 \pi i \alpha}) = 2 \pi^2 \left( \frac{1}{6} - \alpha (1 - \alpha) \right)
\, , \\
&
\ln (1 - {\rm e}^{2 \pi i \alpha}) + \ln (1 - {\rm e}^{- 2 \pi i \alpha}) = 2 \ln \left( 2 \sin (\pi \alpha) \right)
\, ,
\end{align}
one can find the scaling dimension and the normalization constant in Eq.\ \re{TwisAntiTwistCorrAlpha}. The $a = 1$ case has to be studied separately, as one can see the 
appearance of divergences in the expansion \re{GenericaExpantion}. In this regime, we merely quote the result by Kluyver \cite{Kluyver19}
\begin{align}
\ln \prod_{k = 1}^\infty (1 - q^k) 
=
- \frac{\zeta(2)}{\ln \frac{1}{q}} - \frac{1}{2} \ln\ln\frac{1}{q} + \frac{1}{2} \ln (2 \pi) + \frac{1}{24} \ln \frac{1}{q} + \dots
\, ,
\end{align}
which was a predecessor to the Watson estimation \cite{WatsonEst}, i.e., Eq. \re{WatsonEstimation}.


\end{document}